# Sky High 5G:
# New Radio for Air-to-Ground Communications

Xingqin Lin, Anders Furuskär, Olof Liberg, and Sebastian Euler

Ericsson

Contact: xingqin.lin@ericsson.com

*Abstract*— **Today, mobile operators are starting to deploy Fifth-Generation (5G) networks to expand the coverage ubiquity of broadband wireless service. In contrast, in-flight connectivity remains limited and its quality of service does not always meet the expectations. Embracing 5G New Radio (NR) in Air-to-Ground (A2G) communication systems can help narrow the gap between airborne and ground connectivity. In this article, we focus on 5G NR based direct A2G communications. We first provide an overview of the existing A2G systems which are based on earlier generations of mobile technologies. Then we confirm the feasibility of NR A2G systems with a performance study in a range of bands from below 7 GHz to millimeter wave frequencies. The results show that NR A2G systems can provide significantly improved data rates for in-flight connectivity. We also identify the major challenges associated with NR A2G communications, discuss enhancements to counteract the challenges, and point out fruitful avenues for future research.**

## I. INTRODUCTION

The proliferation of smartphones and their use to access the internet have grown to the point where lack of connectivity may result in discomfort and frustration. As current mobile trends develop, there is a clear global demand for ubiquitous wireless connectivity. While mobile broadband service is becoming prevalent on land, in-flight connectivity (IFC) remains limited and its quality of service is often perceived by consumers as poor [1]. Provision of home-quality broadband IFC is an attractive market considering 4.3 billion passengers being carried by airlines in 2018 [2].

IFC can be provided by satellite communication systems and cellular-based direct Air-to-Ground (A2G) communications. In the case of satellite communications, the connectivity between aircraft and ground stations is established by utilizing satellites. The currently available satellite-based IFC solutions mostly operate over the Ku and Ka frequency bands [1][3]. The satellite-based IFC solutions are particularly suitable for intercontinental flights over the ocean, but they usually suffer from limited system capacity and long transmission latencies. Direct A2G communications utilize cellular technology to establish direct connectivity between aircraft and ground stations. The ground stations play a role similar to cellular towers, but their antennas are up-tilted towards the sky. The inter-site distances (ISD) of the ground stations for direct A2G communications are also much greater than their counterparts deployed for terrestrial communications. Compared to the satellite-based solutions, the cellular-based direct A2G solutions have the potential of offering larger system capacity and shorter latencies for IFC and are particularly attractive for short- and medium-haul continental flights and long-haul flights over or near land. But the direct A2G solutions have difficulties in providing connectivity for intercontinental flights over the oceans. Therefore, the satellite-based and cellular-based solutions complement each other, and both are needed to achieve full-scale IFC in the skies.

In this article, we focus on direct A2G communications for IFC. The existing A2G systems for public mobile communications utilize cellular technologies [5]. For example, in the U.S., the Gogo Biz network uses a modified version of the Third-Generation (3G) Code Division Multiple Access (CDMA) 2000 technology to provide IFC. Another example is the European Aviation Network that utilizes Fourth-Generation (4G) Long-Term Evolution (LTE) ground network (in combination with satellite coverage) [6]. As described in more detail in Section II, the existing A2G systems are limited in capacity (typically up to tens of Mbps). They cannot fulfill the vision of providing home-quality broadband to every seat of every aircraft [7].

To provide significantly improved IFC experience for the passengers, the A2G systems will need to be evolved to exploit the Fifth-Generation (5G) wireless access technology, known as New Radio (NR). 5G NR will become a dominant access technology in the next several years, addressing a wide range of use cases from enhanced mobile broadband (eMBB) to ultra-reliable low-latency communications (URLLC) to massive machine type communications (mMTC) [10]. NR features ultra-lean transmission, support for low latency, advanced antenna technologies, and spectrum flexibility including operation in high frequency bands, interworking between high and low frequency bands, and dynamic time-division duplex (TDD) [8]. Embracing NR in A2G systems is expected to provide enhanced performance and vastly improved user experience across a range of flight paths, use cases, and aircraft types. It is worth noting that the 3rd Generation Partnership Project (3GPP) work on NR non-terrestrial networks (NTN) in Release 17 also includes the support of A2G communications [11].

The recent research on A2G communications has mostly focused on low altitude (e.g., a few hundreds of meters) unmanned aerial vehicles [12][13]. For IFC in commercial aircraft typically flying at an altitude between 9.5 km and 12 km, the work of [14] discussed the technical possibilities of enhancing the existing LTE for A2G communications. The work of [15] presented a simulation study for the compatibility of an in-cabin LTE femto cellular system with the current terrestrial LTE systems. In [4], the authors conducted a performance comparison of a 4G A2G network, a 5G A2G network, and a satellite network for IFC. The studied networks were mainly based on the LTE standards, though one of the networks is called "5G A2G network". Preliminary link and system level evaluations of NR A2G systems were carried out in [16]. In contrast, this article provides a more in-depth performance evaluation of NR A2G systems in a range of bands (low, mid, and high). Further, we identify the major challenges associated with NR based direct A2G communications and delve into the detailed NR technical specifications to discuss enhancements to address the challenges. Additionally, we provide an overview of the existing A2G systems and point out some fruitful avenues for future research.

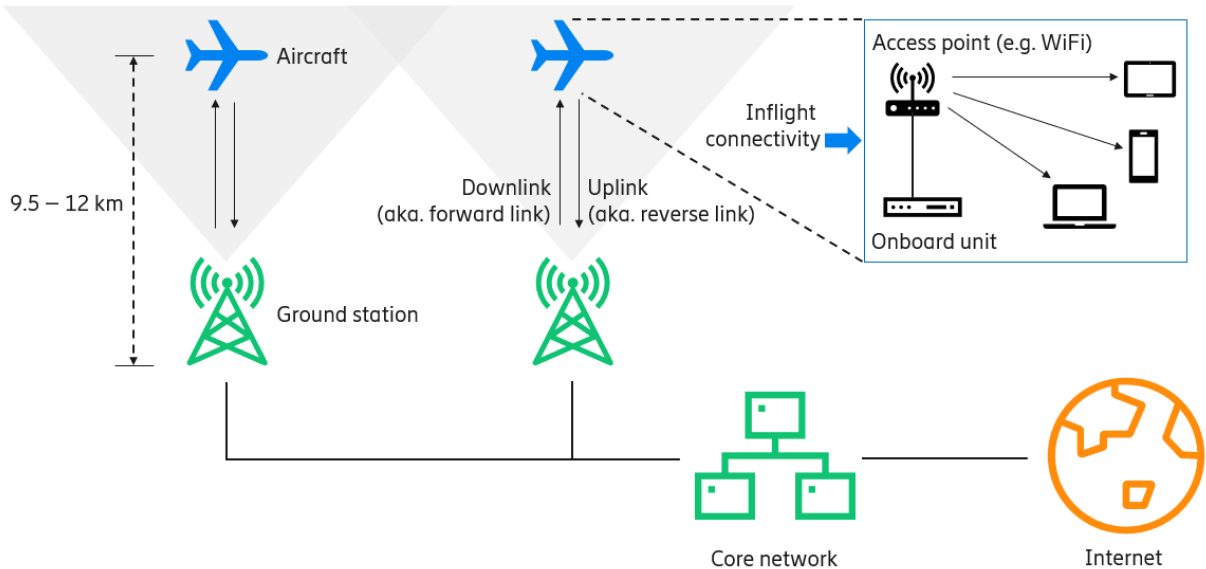

**Figure 1: An illustration of the system architecture for cellular-based direct A2G communications.**

## II. OVERVIEW OF A2G SYSTEMS

In this section, we provide an overview of the existing exemplary A2G systems for public communications [5]. Figure 1 gives an illustration of the system architecture for such systems, which consists of a) cabin access network providing, e.g., WiFi connectivity to end users, b) A2G network equipment onboard aircraft for communicating with ground stations, c) ground radio access network for establishing direct A2G radio links to aircraft, and d) core network for connection management and connectivity to external packet data networks.

### A. A2G Systems in North America

Gogo Biz's A2G network has more than 200 towers in the continental U.S., Alaska, and Canada for providing in-flight WiFi connectivity. It operates in the frequency bands 849-851 MHz (downlink) and 894-896 MHz (uplink). The connection between aircraft and ground stations uses modified Evolution-Data Optimized (EVDO), which is part of the CDMA2000 standard. The maximum total download data rate is 9.8 Mbps. Enhancements were made to handle extended cell size (up to 400 km) and aircraft speed (resulting in an extended range of Doppler shifts and complexities of the airborne handover procedure).

Back in 2011, Qualcomm submitted a petition to the Federal Communications Commission (FCC) on deploying an A2G system (dubbed the *Next-Gen AG* system) operating in the Ku band (14-14.5 GHz) sharing with fixed-satellite service. The proposed system would use between 150 and 250 ground stations scattered around the U.S. to provide an aggregated data rate up to 300 Gbps. The proposed air interface was based on orthogonal frequency-division multiplexing (OFDM), with TDD being the communication mode. The proposed system, however, has not been deployed as of today.

### B. A2G Systems in Europe

Extensive studies have been carried out in Europe to identify suitable frequency band(s) for A2G systems, including 1900-1920 MHz, 2010-2025 MHz, 2400-2483.5 MHz, 3400-3600 MHz, and 5855-5875 MHz.

- A2G system identified in ETSI TR 103 054 [17]: This A2G system was based on the LTE specifications, using paired spectrum of 2×10 MHz for frequency division duplex (FDD) operation. Trials were conducted in Germany within the 2.6 GHz FDD bands. The trial results demonstrated peak data rates of up to 30 Mbps in the downlink and 17 Mbps in the uplink.

- A2G system identified in ETSI TR 101 599 [18]: This A2G system was optimized to operate within the bands 2400-2483.5 MHz and 5855-5875 MHz, utilizing 20 MHz TDD spectrum or 2×10 MHz FDD spectrum. The air interface was based on OFDM. This system featured adaptive beamforming antennas and used four separate phased array antennas at each ground station. Each phased array antenna could generate multiple spatially separated beams to serve the aircraft.

- A2G system identified in ETSI TR 103 108 [19]: The system was designed to operate in the 5855-5875 MHz TDD band and could use 5 MHz or 10 MHz bandwidth. The air interface was Universal Mobile Telecommunications System (UMTS) based on CDMA.

Despite the extensive studies and trials, the commercial deployments of these systems have not yet emerged. In July 2018, the Electronic Communications Committee (ECC) withdrew the previous decision on the harmonized use of A2G systems in the 1900-1920 MHz band. That said, the European Aviation Network, with integrated S-band satellite connection and complementary LTE-based terrestrial network, was launched in 2018. The ground network uses LTE band 65 (2100 MHz) and includes 300 ground stations (up to 75 km cell radius) spread across Europe. The network can provide data rates of up to 75 Mbps in the downlink and 20 Mbps in the uplink.

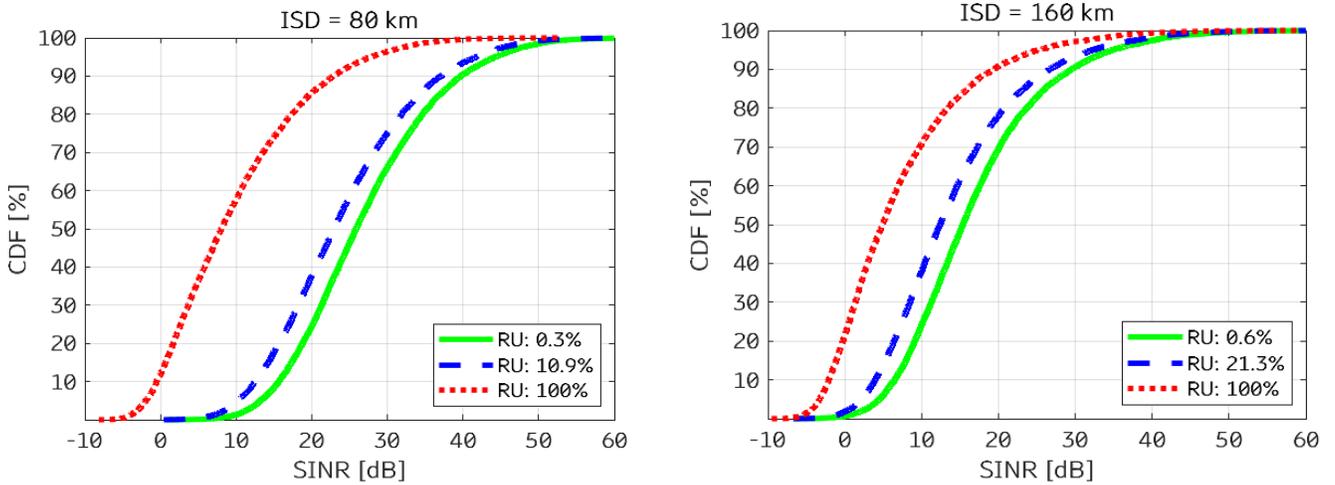

**Figure 2: Downlink SINR distributions under different traffic loads in an NR A2G system at the low band.**

### C. A2G Systems in Asia

In October 2012, the Civil Aviation Administration of China (CAAC) started China's A2G system project. The initial system was based on the synchronous CDMA (SCDMA) specifications and employed TDD as the communication mode [20]. Trials were conducted in the 1785-1805 MHz band. Later, the focus changed to LTE based A2G. Extensive experimental verifications for LTE based A2G in civil aviation applications have been conducted in China in the last few years. It is expected that the A2G system will be commercially available in China in the next few years.

In Japan, several trials were conducted in 2012 to test the performance of a prototype A2G system operating in the 40 GHz frequency range. The system used FDD and employed antenna tracking for proper operation in the millimeter wave frequency. The trial results demonstrated 141.7 Mbps data rate for quadrature phase shift keying (QPSK) modulation and 106.3 Mbps data rate for eight phase shift keying (8PSK) modulation. In 2017, a trial A2G system based on TDD LTE was tested in the very high frequency (VHF) band in Japan. The trial results showed a maximum downlink data rate of 27 Mbps at a flight speed of 430 km/h.

### III. PERFORMANCE STUDY OF NR A2G SYSTEMS

The different A2G systems, as described in Section II, use different cellular technologies and operate in different frequency ranges, from below 7 GHz to millimeter wave frequencies. It will be desirable to adopt a unified standard globally to reap the benefits of economies of scale to provide much enhanced IFC performance and vastly improved user experience. To this end, 5G NR, the next-generation wireless access technology, will be the natural technology choice for future A2G systems. In this section, we present performance evaluation of NR based A2G systems for a range of bands (low, mid, and high) to shed light on the potential of NR for IFC in the 5G era.

### A. NR A2G at Low Band

We first consider an NR A2G system operating at low-band spectrum (below 1 GHz). In the simulation, there are 19 ground stations placed on a hexagonal grid. The NR A2G system uses $2\times10$ MHz FDD spectrum at 700 MHz carrier frequency. Each ground station uses an antenna array with parameters (M, N, P) = (2, 2, 2) to produce a beam, where M denotes the number of rows in the array, N denotes the number of columns in the array, P denotes polarization, and the pattern of each antenna element follows the 3GPP TR 38.901 [21]. These antenna arrays are laid flat facing the sky at a height of 35 m. The aircraft are placed at a height of 12 km, and each has two cross-polarized isotropic antennas. The transmit powers of the ground stations and the aircraft are 80 W and 0.2 W, respectively. The resource utilization (RU) level can indicate the interference level in the network: the higher the RU level, the more the co-channel interference. So, as expected, the SINR becomes worse as the RU level increases. It is also observed that the SINR distributions with 80 km ISD are better than their counterparts with 160 km ISD. For example, the 5-percentile SINR with 80 km ISD at 0.3% RU level is 13.2 dB, which is much higher than the 4.5 dB 5-percentile SINR with 160 km ISD at 0.6% RU level. This is because the signal powers are lower in the larger cells due to larger path loss and smaller antenna gains experienced by the UEs in the network with 160 km ISD that uses a same antenna array as used in the network with 80 km ISD. The SINR difference becomes smaller at the high RU levels where co-channel interference becomes more pronounced. For example, at the 100% RU level, the 5-percentile SINR with 80 km ISD is -2 dB, which is slightly higher than the -3.3 dB 5-percentile SINR with 160 km ISD.

Next, we turn to the throughput performance with 80 km ISD in a single beam setting. Figure 3 shows the downlink and uplink throughput distributions at different RU levels at the low band. Since in-flight traffic is typically downlink heavy, we focus on examining the throughput performance at high load in the downlink and at low load in the uplink. At the RU level of 79%, the 5-, 50-, and 99-percentile downlink throughput values are 1.6 Mbps, 8.5 Mbps, and 32.2 Mbps, respectively. At the RU level of 1.7%, the 5-, 50-, and 90-percentile uplink throughput values are 0.35 Mbps, 7.6 Mbps, and 19.2 Mbps, respectively. These throughput values appear to be on par with the data rates offered by the LTE-based European Aviation Network (up to 75 Mbps in the downlink and 20 Mbps in the uplink).

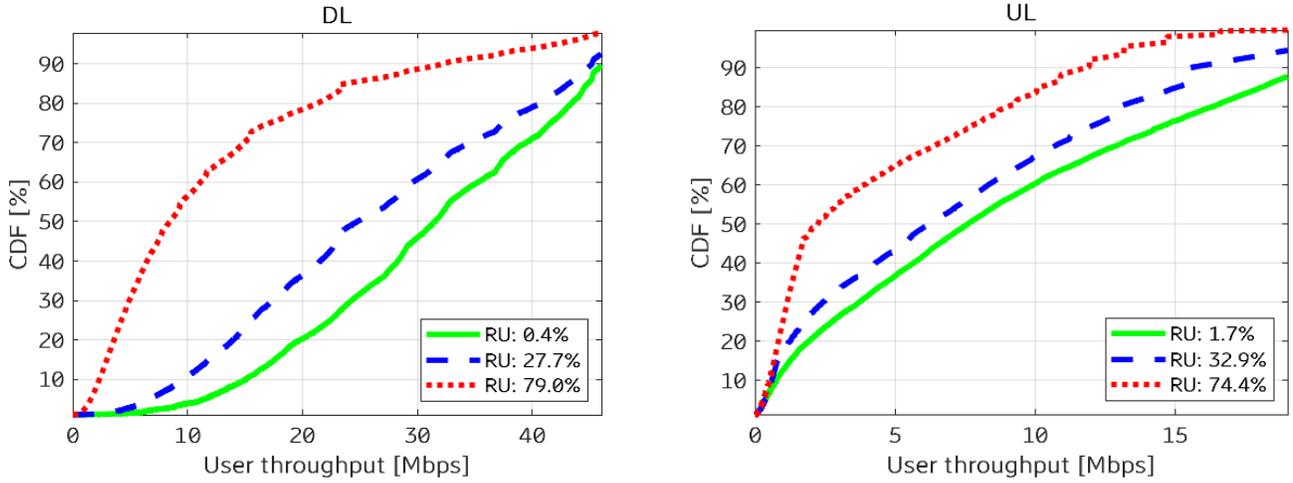

**Figure 3: Throughput distributions under different traffic loads in an NR A2G system at the low band.**

### B. NR A2G at Mid- and High Band

NR features spectrum flexibility and supports operation in the spectrum ranging from sub-1 GHz to millimeter wave bands. To further explore the potential of NR based A2G systems, we next turn to NR A2G systems operating in mid-band (1-7 GHz) and high band (millimeter wave frequencies).

The NR A2G system at the mid-band spectrum uses 2×100 MHz FDD spectrum at 3.5 GHz carrier frequency and an antenna array with parameters (M, N, P) = (4, 4, 2) at the ground station. One so produced beam covers 1/4 of the area covered by one beam at the low band. So, if 80 km ISD is kept in the mid-band, each ground station should produce 4 beams to cover a cell. Figure 4 shows the downlink and uplink throughput distributions at different RU levels at the mid-band. At the RU level of 82.3% in the downlink, the 5-, 50-, and 99-percentile throughput values are 5.9 Mbps, 40.6 Mbps, and 175.6 Mbps, respectively. At the RU level of 2.3% in the uplink, the 5-, 50-, and 99-percentile uplink throughput values are 0.72 Mbps, 19.3 Mbps, and 112.0 Mbps, respectively. The highest downlink and uplink throughput values are 454.9 Mbps and 197.5 Mbps, respectively. We can see that by exploiting the large bandwidth in the mid-band, this NR A2G system offers much higher throughput values than its counterpart in the low band.

The NR A2G system at the high band uses 2×400 MHz FDD spectrum at 28 GHz carrier frequency and an antenna array with parameters (M, N, P) = (8, 8, 2) at the ground station. One so produced beam covers 1/64 of the area covered by one beam at the low band. So, if 80 km ISD is kept in the high band, each ground station should produce 64 beams to cover a cell. Figure 5 shows the downlink and uplink throughput distributions at different RU levels at the high band. The highest downlink and uplink throughput values are 1.5 Gbps and 563.9 Mbps, respectively. This NR A2G system is capable of providing Gbps links to the aircraft.

## IV. POTENTAL ENHANCEMENTS FOR NR A2G SYSTEMS

In the previous section, we have presented a performance study of NR based A2G systems for a range of frequency bands, illustrating the potential of NR for IFC. Though the inherent flexibility of NR allows it to be used to support A2G communications, NR has been designed mainly targeting terrestrial mobile communications. In this section, we discuss performance enhancing solutions to optimize NR connectivity to provide further improved performance for IFC.

### A. Large Cells

A2G systems feature large cells to limit network deployment cost for serving sparsely scattered aircraft in the sky. Typical ISD in A2G systems is expected to range from 80-200 km. In some cases, a larger cell size may be needed, for example, to enable offshore aircraft flying close to coast to communicate with nearest ground stations. To accommodate these cases, we consider a maximum cell radius of 300 km as an appropriate design target for NR A2G systems.

Supporting NR A2G systems with up to 300 km cell radius across a range of bands requires revisiting the many timing relationships defined in NR specifications. For example, timing advance is used at the User Equipment (UE) to adjust uplink frame timing relative to downlink frame timing. The required timing advance for a UE is roughly equal to the round-trip delay between the UE and the serving 5G NodeB (gNB), e.g., up to ~2 ms for an A2G system with cell radii up to 300 km. During a random-access procedure, the gNB estimates the required timing advance value by processing the received random-access preamble and sends the value to the UE in a random-access response message. The maximum timing advance applied during initial access in NR is equal to $2^{-\mu} \times 2$ ms for subcarrier spacing values of $2^{\mu} \times 15$ kHz, where $\mu = 0, 1, 2, 3, 4$. So, except for the 15 kHz subcarrier spacing, the timing advance value range is not sufficient to support NR A2G systems across a range of bands and, thus, would need to be extended.

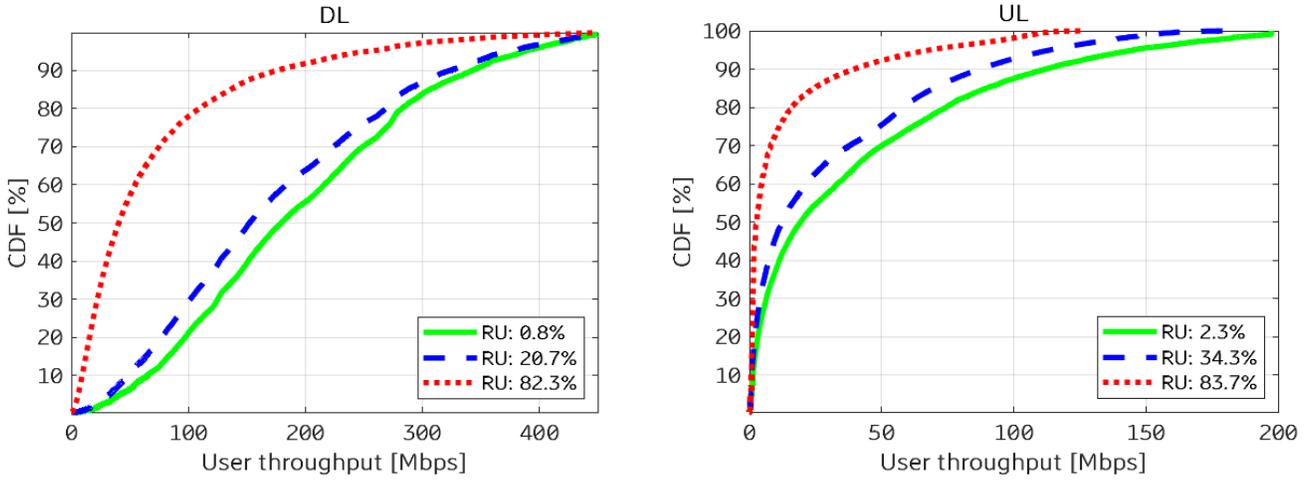

**Figure 4: Throughput distributions under different traffic loads in an NR A2G system at the mid-band.**

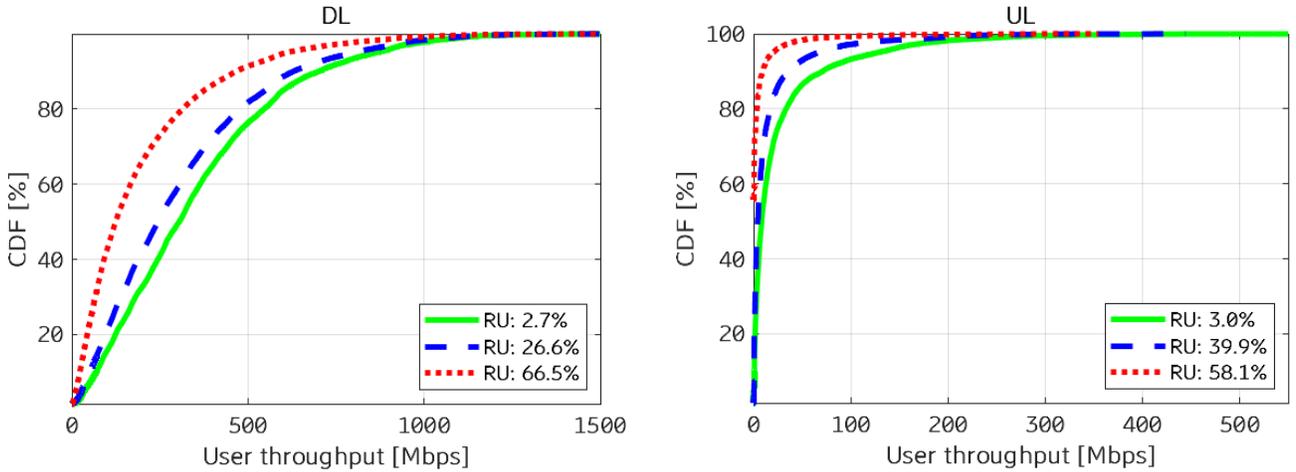

**Figure 5: Throughput distributions under different traffic loads in an NR A2G system at the high band.**

In case of TDD based NR A2G systems, a guard period is required to isolate downlink slots from uplink slots. The maximum guard period would be 2 ms to support 300 km cell radius. To limit the overhead of guard period to be no more than 10%, the TDD period would need to be at least 20 ms. Consider, for example, the 30 kHz subcarrier spacing with 0.5 ms slot duration, which is a popular design choice for NR deployments in the mid-band spectrum. In this case, the 2 ms guard period translates into 4 guard period slots. Assuming 20 ms TDD period, one example TDD frame structure could include 24 downlink slots, 4 guard period slots, and 12 uplink slots in one period, where the number of downlink slots is twice of the number of uplink slots to adapt to the downlink heavy in-flight traffic. For a physical downlink shared channel (PDSCH) reception ending in the slot $n$, a UE may need to transmit hybrid automatic repeat request (HARQ) feedback in the slot $n + k_1$, where $k_1$ indicates the slot offset between the PDSCH ending slot and the slot for HARQ transmission. The maximum possible value of $k_1$ in NR is 15, while the aforementioned example TDD structure with 30 kHz subcarrier spacing would require the value of $k_1$ to be configurable up to 39.

### B. High Mobility

Commercial aircraft cruise at about 740–930 km/h. We consider a maximum UE speed of 1200 km/h to be the design target for NR A2G systems.

The high UE speeds in A2G systems result in pronounced Doppler effects. At the speed of 1200 km/h, a UE would experience Doppler shifts of up to about $\pm 1.11$ ppm in the downlink, i.e., about $\pm 0.78$ kHz, $\pm 3.89$ kHz, and $\pm 31.08$ kHz at the carrier frequencies of 700 MHz, 3.5 GHz, and 28 GHz, respectively. Handling such high Doppler shifts may need new UE performance requirements for NR A2G systems. The Doppler shifts in the uplink would be about twice of the Doppler shifts in the downlink. The severe Doppler effects may cause inter-carrier interference in the uplink. To mitigate the inter-carrier interference, a gNB may schedule different UEs in different frequencies with sufficient guard spectrum in-between. However, this is not a spectrally efficient solution. An enhancement which can help the transmissions from different UEs in a cell to be frequency aligned at the gNB would be more desirable. This can be achieved by applying different frequency adjustment values at different UEs in the uplink to compensate for their different Doppler shifts.

The high aircraft speeds in A2G systems also bring in challenges for antenna beam tracking, though the NR channels, signals, and procedures have been designed to support beamforming. The volumes of cone-shaped beams become larger due to the larger ISD values in A2G systems, increasing the probability of intersecting beams and potentially larger inter-beam interference. Nonetheless, the A2G channels are line-of-sight (LOS) dominated. The beamforming can exploit the location information of aircraft which usually have fixed flight routes and stable mobility patterns. The ground stations may obtain the location information of the aircraft by listening to, for example, the Automatic Dependent Surveillance Broadcasting (ADS-B) signal that includes the position, speed, and altitude of aircraft. The beamforming at the UE may be further facilitated if the location information of the ground stations is made available at the UE side. The location information assisted beamforming can help improve signal strength and reduce interference. The information transmitted in ADS-B may also be utilized for Doppler estimation and compensation.

Despite the high UE speeds in A2G systems, handover events are not expected to be frequent due to the large cell sizes. For example, it would take a few minutes for an aircraft cruising at 1200 km/h to traverse through a cell with 50 km radius. This handover rate is lower than the handover rate that a high-speed train would experience in a terrestrial network where cells are of much smaller sizes. As NR is capable of serving high-speed trains, mobility management is not expected to be challenging for NR A2G systems. Certain enhancements may be considered to further improve the mobility performance in NR A2G systems. For example, the triggering of measurement reporting and/or conditional handover may be made dependent on aircraft UE location.

### C. Coexistence with Terrestrial and Satellite Systems

As indicated in the performance study presented in Section III, large bandwidths are crucial for providing high data rates in NR A2G systems. Securing harmonized large bandwidths dedicated to the A2G systems might be challenging. An alternative could be that mobile operators reuse their terrestrial spectrum for A2G services if permitted by regulation.

Using the same spectrum for both terrestrial 5G network and A2G network requires careful deployment planning to ensure that mutual interference between the terrestrial 5G and A2G networks is acceptable. Such deployment coordination may be easier for the upper portion of mid-band spectrum and high-band spectrum, which are typically used for local deployment for capacity enhancement in terrestrial networks. Thus, the terrestrial gNBs may be geographically separated from the A2G gNBs which can be placed in the remote areas such as on remote mountains. Nonetheless, there would likely be many complications obstructing co-channel deployment of terrestrial 5G and A2G networks, for which radio frequency (RF) requirements of base stations for A2G and aircraft UEs would need to be studied.

Note that the A2G ground stations are deployed to serve aircraft UEs, and they do not aim to directly serve the passengers' UEs. The use of mobile phones, tablets, and laptops during the flight typically requires the devices to be switched to airplane mode. The devices without airplane mode switched on may continuously try to access the A2G cells, draining the batteries of the devices and causing vain access loads to the A2G system. To prevent this, NR A2G would need an access control mechanism to give access exclusively to aircraft.

One might also consider reusing the satellite spectrum for A2G services, similar to Qualcomm's petition on deploying an A2G system operating in the Ku band used for fixed-satellite service. The incumbent satellite services should be protected from interference from such A2G systems. Regulation and interference coordination issues in this case appear to be even more challenging than the co-channel deployment of terrestrial 5G and A2G networks.

## V. CONCLUSIONS AND RESEARCH DIRECTIONS

Wireless broadband connectivity is becoming ubiquitous, and the sky should not impose a limit on that. The quality of the current in-flight connectivity service is unsatisfactory. It is of importance to develop solutions to provide true broadband connectivity in the cabin. 5G NR will become the new normal in the next several years. The existing A2G systems are based on earlier generations of mobile technologies. This article has presented a performance study of NR A2G systems in a range of bands. The results show that embracing 5G NR in the A2G systems has the potential of providing enhanced performance and vastly improved user experience. This article has also identified the major challenges and discussed enhancements for NR A2G systems.

Making sky high 5G broadband connectivity a reality requires breaking down many barriers along the road. We conclude by pointing out some fruitful avenues for future research.

*Aircraft UE beamforming:* The performance study of NR A2G systems in this article has assumed beamforming at the ground stations. Beamforming at the aircraft UEs has the potential of further improving the system performance. It is important to make the antenna design and operation compatible with aircraft engineering and operations.

*Interference management for A2G systems*: Beamforming and beam-steering techniques deliver a directional signal to the aircraft. The beams may intersect in the skies and cause mutual interference. Coordinating resource allocation and beam management for interference mitigation in A2G systems is an interesting research problem. Besides, coexistence studies between A2G systems and other terrestrial/satellite systems are of high interest to ensure that the systems do not cause harmful interference to each other.

*Prototyping of NR A2G systems*: For further understanding the potential and challenges of NR A2G systems, it is important to develop early prototypes and collect feedback. The prototypes may help identify potential shortcomings in the NR specifications for A2G systems. The prompt feedback would facilitate the adoption of appropriate enhancements in the NR standards.

*Integrated terrestrial 5G, A2G, and satellite networks*: The future of connectivity will be seamless, regardless of where you are. True seamless connectivity will need a network of networks that integrate terrestrial 5G, A2G, and satellite networks, among others. Designing and managing the network of networks to provide transparent service continuity to users is challenging, but important.

*Spectrum, regulation, and business models for A2G systems*: Harmonized spectrum allocations and unified regulatory frameworks across national borders are key to a significant uptake of A2G systems. The right business models should also be in place to help achieve sufficient market scale for A2G systems. It is vital to develop an agreed set of international standards to build a successful A2G ecosystem to achieve seamless in-flight broadband connectivity.